\documentclass[12pt]{article}        

\usepackage{amsmath}
\usepackage{amssymb}
\usepackage{euscript}

\usepackage{epsfig}

\usepackage{cite}

\usepackage{epic}
\usepackage{fancybox}

\def\Order#1{${\cal O}(#1$)}

\def\Oceex#1{${\cal O}(#1)_{_{\rm CEEX}}$}

\newcommand{\Reu}{\EuScript{R}}

\newcommand{\Meu}{\EuScript{M}}

\newcommand{\Bmf}{\mathfrak{B}}

\newcommand{\Mmf}{\mathfrak{M}}

\newcommand{\umf}{\mathfrak{u}}
\newcommand{\Mcal}{{\cal M}}

\newcommand{\sfac}{\mathfrak{s}}
\begin{document}                     

\allowdisplaybreaks

\begin{titlepage}


\begin{flushright}
CERN-TH-98-253\\
UTHEP-98-0801
\end{flushright}

\vspace{1mm}
\begin{center}
{\Large\bf
  Coherent Exclusive Exponentiation CEEX: \\
  the Case of the Resonant $e^+e^-$ Collision
}
\end{center}
\vspace{3mm}

\begin{center}
{\bf S. Jadach$^{a,b}$},
{\bf B.F.L. Ward$^{a,c,d}$}
{\em and}
{\bf Z. W\c as$^{a,b}$} \\

\vspace{1mm}
{\em $^a$CERN, Theory Division, CH-1211 Geneva 23, Switzerland,}\\
{\em $^b$Institute of Nuclear Physics,
  ul. Kawiory 26a, Krak\'ow, Poland,}\\
{\em $^c$Department of Physics and Astronomy,\\
  The University of Tennessee, Knoxville, TN 37996-1200,\\
  $^d$SLAC, Stanford University, Stanford, CA 94309} 
\end{center}

\vspace{10mm}
\begin{abstract}
We present the first-order coherent exclusive exponentiation (CEEX) scheme, 
with the full control over spin polarization for all fermions.
In particular it is applicable to difficult case of narrow resonances.
The resulting spin amplitudes and the differential distributions are given in a form
ready for their implementation in the Monte Carlo event generator.
The initial-final state interferences are under control.
The way is open to the use of the exact amplitudes for two and more hard photons,
using Weyl-spinor techniques,
without giving up the advantages of the exclusive exponentiation,
of the Yennie-Frautschi-Suura type.
\end{abstract}

\begin{center}
{\it Submitted to Phys. Lett. B}
\end{center}

\vspace{1mm}
\begin{flushleft}
{\bf CERN-TH/98-253\\
  UTHEP-98-0801\\
     August,~1998}
\end{flushleft}

\footnoterule
\noindent
{\footnotesize
\begin{itemize}
\item[${\dagger}$]
Work supported in part by 
the US DoE contract DE-FG05-91ER40627 and DE-AC03-76SF00515,
Polish Government grants 
KBN 2P03B08414, 
KBN 2P03B14715, 
Maria Sk\l{}odowska-Curie Fund II PAA/DOE-97-316,
and Polish-French Collaboration within IN2P3.
\end{itemize}
}

\end{titlepage}





\section{\normalsize Introduction: What is the problem?}

$\EuScript{}$ 
The problem addressed in this work is: 
How to describe,
consistently in the process $e^+e^-\to f\bar{f}$,
the coherent emission of {\em initial} state radiation (ISR) 
and {\em final} state radiation (FSR) of {\em soft} and {\em hard} photons,
providing for cancellations of infrared (IR) divergences from real and virtual
photon emission to infinite perturbative order (exponentiation),
at the level of completely {\em exclusive} multiphoton differential distributions,
i.e. in the form suitable for implementation in Monte Carlo (MC) event generators?
In addition we are looking for the solution 
that is friendliest to narrow $s$-channel resonances.

This work is firmly rooted in the work of Yennie, Frautschi  and Suura (YFS)
on QED exponentiation \cite{yfs:1961}
and its further developments in refs.~\cite{yfs1:1988,yfs2:1990,bhlumi2:1992}.
The present work definitely goes beyond the scope of these previous papers --
the main difference is the consequent use of spin amplitudes in the exponentiation.
Our work is close in spirit, although not in technical details,
to seminal papers of Greco et~al. \cite{greco:1975,greco:1980}
on QED exponentiation for narrow resonances.
However, it should be stressed that, contrary 
to refs.~\cite{greco:1975,greco:1980}, all our differential
multiphoton distributions are completely exclusive (important for MC implementation)
and we do include hard photons completely and systematically.
In this context, the work of ref.~\cite{bhwide:1997} should also be mentioned.
It implements QED interferences among $e^+$ and $e^-$ fermion lines, the analog of
the ISR--FSR interferences, for the first time in the exclusive exponentiation.
It does not, however, use spin amplitudes for exponentiation as consequently
as does the present work;  it is also rather strongly limited to exact first order
exponentiation in the YFS framework.
It is sort of half-way between the present work and the older ones 
of refs.~\cite{yfs1:1988,yfs2:1990}.
At the technical level,  the methods used here for the construction of the spin amplitudes
are essentially those%
\footnote{ We have evaluated several techniques based on Weyl-spinor techniques
  and we concluded that the technique of KS is best suited 
  for our needs (exponentiation).}
of Kleiss and Stirling (KS) \cite{KleissStirling:1985,KleissStirling:1986},
with the important supplement of ref.~\cite{gps:1998},
providing for total control of complex phases and/or fermion spin quantization frames.
The MC implementation of the present work will soon be available
\cite{KK:1998} and it will replace two MC programs:
KORALB \cite{KORALB:1985}, where fermion spin polarizations are implemented exactly,
but there is no exponentiation,
and KORALZ \cite{KORALZ_4.0}, where exponentiation is included,
but the treatment of spin effects is simplified%
\footnote{ The ISR--FSR interference is also neglected in KORALZ,
  in the main mode with the exponentiation switched on.}.

The present work is essential for any present
experiments in $e^+e^-$ colliders and future $e^+e^-$ and $\mu^+\mu^-$ colliders,
where the most important new features for data analysis
will be inclusion of ISR--FSR interferences and
(in the next step) the exact matrix element for emission of 2 and 3 hard
photons, in the presence of many additional soft ones.

\section{\normalsize Basic KS/GPS spinors and photon polarizations}

The arbitrary massless spinor $u_\lambda(p)$
of momentum $p$ and chirality $\lambda$ is defined according to KS methods
\cite{KleissStirling:1985,KleissStirling:1986}.
In the following we follow closely the notation of ref.~\cite{gps:1998}
(in particular we also use $\zeta=\zeta_{\downarrow}$).
In the above framework every spinor is transformed out of the two 
{\em constant basic} spinors
$\umf_\lambda(\zeta)$, of opposite chirality $\lambda=\pm$, as follows
\begin{equation}
\label{def-massless}
  u_\lambda(p) 
     = {1\over \sqrt{ 2p\cdot \zeta}} \not\!p  \umf_{-\lambda}(\zeta),\quad
     \umf_+(\zeta)=\not\!\eta  \umf_-(\zeta),\quad
     \eta^2=-1,\quad
     (\eta\zeta)=0.
\end{equation}
The usual relations hold:
 $\not\!\zeta \umf_\lambda(\zeta) =0$,
 $\omega_\lambda \umf_\lambda(\zeta) =\umf_\lambda(\zeta)$,
 $\umf_\lambda(\zeta) \bar{\umf}_\lambda(\zeta) = \not\!\zeta \omega_\lambda $,
 $\not\!p u_\lambda(p) =0$,
 $\omega_\lambda u_\lambda(p) =u_\lambda(p)$,
 $u_\lambda(p) \bar{u}_\lambda(p) = \not\!\!p \omega_\lambda$, where
 $\omega_\lambda = {1\over 2} (1+\lambda \gamma_5)$.
Spinors for the massive particle with four-momentum $p$ (with $p^2=m^2$)
and spin projection $\lambda/2$ are defined
in terms of massless spinors
\begin{equation}
\label{def-massive2}
u(p,\lambda)
        = u_\lambda(p_\zeta)    +{m\over \sqrt{2p\zeta}} \umf_{-\lambda}(\zeta),\qquad
v(p,\lambda)
        = u_{-\lambda}(p_\zeta) -{m\over \sqrt{2p\zeta}} \umf_{\lambda}(\zeta),
\end{equation}
where 
$  p_\zeta  \equiv \hat{p}  \equiv p - \zeta\; m^2/(2\zeta p)$
is the light-cone projection ($ p_\zeta^2 =0 $) 
of the $p$ obtained with the help of the constant auxiliary vector $\zeta$.

The above definition is supplemented in ref.~\cite{gps:1998} with the precise prescription
on spin quantization axes, translation from spin amplitudes to density matrices
(also in vector notation) and the methodology of connecting production and decay
for unstable fermions.
We collectively call these rules global positioning of spin (GPS).
Thanks to these we are able to easily introduce polarizations for beams
and implement polarization effects for final fermion decays 
(of $\tau$ leptons, t-quarks), for the first time
also in the presence of emission of many ISR and FSR photons!

The GPS rules determining spin quantization frame
for $u(p,\pm)$ and $v(p,\pm)$ of eq.~(\ref{def-massive2})
are summarized as follows:
(a)
In the rest frame of the fermion, take the $z$-axis along $-\vec{\zeta}$.
(b)
Place the $x$-axis in the plane defined by the $z$-axis from the previous point
and the vector $\vec{\eta}$, in the same half-plane as $\vec{\eta}$.
(c)
With the $y$-axis, complete the right-handed system of coordinates.
The rest frame defined in this way we call the GPS frame of the particular fermion.
See ref.~\cite{gps:1998} for more details.
In the following we shall assume that polarization vectors of beams
and of outgoing fermions are defined in their corresponding GPS frames.

The inner product of the two massless spinors is defined as follows
  \begin{equation}
        s_{+}(p_1,p_2) \equiv \bar{u}_{+}(p_1) u_{-}(p_2),\quad
        s_{-}(p_1,p_2) \equiv \bar{u}_{-}(p_1) u_{+}(p_2) 
                       = -( s_{+}(p_1,p_2))^*.
  \end{equation}
The above inner product can be evaluated
using the Kleiss-Stirling expression
\begin{equation}
  s_+(p,q) = 2\; (2p\zeta)^{-1/2}\; (2q\zeta)^{-1/2}\; 
  \left[ (p\zeta)(q\eta) - (p\eta)(q\zeta) 
    -i\epsilon_{\mu\nu\rho\sigma} \zeta^\mu \eta^\nu p^\rho q^\sigma 
  \right]
\end{equation}
in any reference frame.
In particular, in the laboratory frame we typically use
$\zeta=(1,1,0,0)$ and $\eta=(0,0,1,0)$, which leads to the following ``massless''
inner product
\begin{equation}
  s_+(p,q) = -(q^2 + iq^3) \sqrt{ (p^0 -p^1)/( q^0-q^1) }
             +(p^2 + ip^3) \sqrt{ (q^0 -q^1)/( p^0-p^1) }.
\end{equation}
Equation~(\ref{def-massive2}) immediately
provides us also with the {\em inner product} for massive spinors
\begin{equation}
  \begin{split}
    & \bar{u}(p_1,\lambda_1)  u(p_2,\lambda_2)
                =S(p_1,m_1,\lambda_1,  p_2,m_2,\lambda_2), \\
    & \bar{u}(p_1,\lambda_1)  v(p_2,\lambda_2)
                =S(p_1,m_1,\lambda_1,  p_2,-m_2,-\lambda_2), \\
    & \bar{v}(p_1,\lambda_1)  u(p_2,\lambda_2)
                =S(p_1,-m_1,-\lambda_1,  p_2,m_2,\lambda_2), \\
    & \bar{v}(p_1,\lambda_1)  v(p_2,\lambda_2)
                =S(p_1,-m_1,-\lambda_1,  p_2,-m_2,-\lambda_2), \\
  \end{split}
\end{equation}
where 
\begin{equation}
\label{inner-massive}
  S(p_1,m_1,\lambda_1,  p_2,m_2,\lambda_2)
  = \delta_{\lambda_1,-\lambda_2} s_{\lambda_1}({p_1}_\zeta, {p_2}_\zeta)
   +\delta_{\lambda_1, \lambda_2} 
  \left(
     m_1 \sqrt{ {2\zeta p_2 \over 2\zeta p_1}  }
    +m_2 \sqrt{ {2\zeta p_1 \over 2\zeta p_2}  }
  \right).
\end{equation}
In our spinor algebra we shall exploit the completeness relations
\begin{equation}
  \begin{split}
    &\not\!{p}+m = \sum_\lambda u(p,\lambda) \bar{u}(p,\lambda),\quad
     \not\!{p}-m = \sum_\lambda v(p,\lambda) \bar{v}(p,\lambda),\\
    &\not\!{k}   = \sum_\lambda u(k,\lambda) \bar{u}(k,\lambda),\quad k^2=0.
  \end{split}
\end{equation}

For a circularly polarized photon with four-momentum $k$
and helicity $\sigma=\pm 1$ we adopt the KS choice
(see also ref.~\cite{Beijing:1987}) of polarization vector%
\footnote{ Contrary to other papers on Weyl spinor techniques
  \protect\cite{KleissStirling:1985,CALKUL}
  we keep here the explicitly complex conjugation in $\epsilon$. This conjugation
  is cancelled by another conjugation following from Feynman rules, but only
  for outgoing photons, not for beam photon, as in the Compton process, 
  see ref.~\protect\cite{Stuart:1989}.
  }
\begin{equation}
\label{phot-pol}
  (\epsilon^\mu_\sigma(\beta))^*
     ={\bar{u}_\sigma(k) \gamma^\mu u_\sigma(\beta)
       \over \sqrt{2}\; \bar{u}_{-\sigma}(k) u_\sigma(\beta)},\quad
  (\epsilon^\mu_\sigma(\zeta))^*
     ={\bar{u}_\sigma(k) \gamma^\mu \umf_\sigma(\zeta)
       \over \sqrt{2}\; \bar{u}_{-\sigma}(k) \umf_\sigma(\zeta)},
\end{equation}
where $\beta$ is an arbitrary light-like four-vector $\beta^2=0$.
The second choice with $\umf_\sigma(\zeta)$ 
(not exploited in \cite{KleissStirling:1985})
often leads  to simplifications in the resulting photon emission amplitudes.
Using the Chisholm identity%
\footnote{ For $\beta=\zeta$ the identity is slightly different 
  because of the additional minus sign in the ``line-reversal'' rule, i.e. 
  $ \bar{u}_\sigma(k) \gamma^\mu \umf_\sigma(\zeta) = 
  - \bar{\umf}_{-\sigma}(\zeta) \gamma^\mu u_{-\sigma}(k)$,
  in contrast to the usual
  $ \bar{u}_\sigma(k) \gamma^\mu u_\sigma(\beta) = 
  + \bar{u}_{-\sigma}(\beta) \gamma^\mu u_{-\sigma}(k).$}
\begin{align}
  \label{Chisholm}
    \bar{u}_\sigma(k) \gamma_\mu u_\sigma(\beta)\; \gamma^\mu
   &= 2    u_\sigma(\beta)\;   \bar{u}_\sigma(k)
    + 2    u_{-\sigma}(k)\;    \bar{u}_{-\sigma}(\beta),\\
  \label{Chisholm2}
    \bar{u}_\sigma(k) \gamma_\mu \umf_\sigma(\zeta)\; \gamma^\mu
   &= 2 \umf_\sigma(\zeta)\;    \bar{u}_\sigma(k)      
    - 2 u_{-\sigma}(k)\;    \bar{\umf}_{-\sigma}(\zeta),
\end{align}
we get two useful expressions, equivalent to eq.~(\ref{phot-pol}):
\begin{equation}
  \begin{split}
   &({\not\!\epsilon}_\sigma (k,\beta) )^*
    = {\sqrt{2} \over \bar{u}_{-\sigma}(k) u_\sigma(\beta)}
    \left[ u_\sigma(\beta)  \bar{u}_\sigma(k) 
      +u_{-\sigma}(k)   \bar{u}_{-\sigma}(\beta)
    \right]\\
   &({\not\!\epsilon}_\sigma (k,\zeta) )^*
    = {\sqrt{2} \over \sqrt{2\zeta k} }
    \left[ \umf_\sigma(\zeta)  \bar{u}_\sigma(k) 
      -u_{-\sigma}(k)   \bar{\umf}_{-\sigma}(\zeta)
    \right].\\
  \end{split}
\end{equation}

In the evaluation of photon emission
spin amplitudes we shall use the following important building block --
the elements of the ``transition matrices''  $U$ and $V$  defined as follows
\begin{equation}
  \begin{split}
    \label{transition-defs}
   &\bar{u}(p_1,\lambda_1) 
         \not\!{\epsilon}^\star_\sigma(k,\beta)\;
    u(p_2,\lambda_2)
  = U\left( \hbox{}^k_\sigma \right)\!
     \left[ \hbox{}^{p_1}_{\lambda_1} \hbox{}^{p_2}_{\lambda_2} \right]
  = U^\sigma_{\lambda_1,\lambda_2} (k,p_1,m_1,p_2,m_2),\\
   &\bar{v}(p_1,\lambda_1) 
         \not\!{\epsilon}^\star_\sigma(k,\zeta)\;
    v(p_2,\lambda_2)
  = V\left( \hbox{}^k_\sigma \right)\!
     \left[ \hbox{}^{p_1}_{\lambda_1} \hbox{}^{p_2}_{\lambda_2} \right]
  = V^\sigma_{\lambda_1,\lambda_2} (k,p_1,m_1,p_2,m_2).\\
  \end{split}
\end{equation}
In the case of $\umf_\sigma(\zeta)$ 
the above transition matrices are rather simple%
\footnote{
  Our $U$ and $V$ matrices are not the same as
  the $M$-matrices of ref.~\protect\cite{KleissStirling:1986}, but rather
  products of several of those.
  }:
\begin{equation}
  \begin{align}
  \label{UVsimple}
   &U^+(k,p_1,m_1,p_2,m_2) =\sqrt{2} 
   \begin{bmatrix} 
        \sqrt{ {2\zeta p_2 \over 2\zeta k} } s_+(k,\hat{p_1}),
      & 0 \\ 
        m_2\sqrt{ {2\zeta p_1 \over 2\zeta p_2} }
       -m_1\sqrt{ {2\zeta p_2 \over 2\zeta p_1} },
      & \sqrt{ {2\zeta p_1 \over 2\zeta k} } s_+(k,\hat{p_2})\\
    \end{bmatrix},\\
    \label{UVsimple2}
   &      U^-_{ \lambda_1, \lambda_2}(k,p_1, m_1,p_2, m_2) 
  =\left[-U^+_{ \lambda_2, \lambda_1}(k,p_2, m_2,p_1, m_1)  \right]^*,
\\
\label{UVsimple3}
   &V^\sigma_{ \lambda_1, \lambda_2}(k,p_1, m_1,p_2, m_2) 
   =U^\sigma_{-\lambda_1,-\lambda_2}(k,p_1,-m_1,p_2,-m_2).
  \end{align}
\end{equation}
The more general case with $u_\sigma(\beta)$ looks a little bit more complicated:
\begin{equation}
  \label{UVgeneral}
  \begin{split}
   &U^+(k,p_1,m_1,p_2,m_2) =
   {\sqrt{2} \over s_-(k,\beta)}\times \\
   &
    \begin{bmatrix} 
         s_+(\hat{p}_1,k) s_-(\beta,\hat{p}_2)
        +m_1m_2\sqrt{ {2\zeta \beta \over 2\zeta p_1}
                      {2\zeta k     \over 2\zeta p_2} },
      &  m_1 \sqrt{ {2\zeta \beta \over 2\zeta p_1} } s_+(k,\hat{p}_2)
        +m_2 \sqrt{ {2\zeta \beta \over 2\zeta p_2} } s_+(\hat{p}_1,k)\\ 
         m_1 \sqrt{ {2\zeta k \over 2\zeta p_1} } s_-(\beta,\hat{p}_2)
        +m_2 \sqrt{ {2\zeta k \over 2\zeta p_2} } s_-(\hat{p}_1,\beta),
      &  s_-(\hat{p}_1,\beta) s_+(k,\hat{p}_2)
        +m_1m_2\sqrt{ {2\zeta \beta \over 2\zeta p_1}
                      {2\zeta k     \over 2\zeta p_2} }\\
    \end{bmatrix},\\
  \end{split}
\end{equation}
with the same relations (\ref{UVsimple2}) and (\ref{UVsimple3}).
In the above the following numbering of elements in matrices 
$U$ and $V$ is adopted
\begin{equation}
  \{ (\lambda_1,\lambda_2) \} = 
    \begin{bmatrix}
      (++) & (+-) \\
      (-+) & (--) \\
    \end{bmatrix}.
\end{equation}
When analysing the soft real photon limit
we shall exploit the following important {\em diagonality} property%
\footnote{
  Let us also keep in mind the relation
  $ b_{-\sigma}(k,p)= -(b_\sigma(k,p))^* $, which can
  save time in the numerical calculations.}
\begin{equation}
  \begin{align}
    \label{diagonality}
&   U\left(  \hbox{}^k_\sigma  \right)\!
     \left[  \hbox{}^{p}_{\lambda_1}
             \hbox{}^{p}_{\lambda_2}
           \right]
    = 
    V\left(  \hbox{}^k_\sigma  \right)\!
     \left[  \hbox{}^{p}_{\lambda_1}
             \hbox{}^{p}_{\lambda_2}
           \right]
    = b_\sigma(k,p)\; \delta_{\lambda_1 \lambda_2},\\
&   \label{b-sigma}
    b_\sigma(k,p)
    =  \sqrt{2}\; { \bar{u}_\sigma(k) \not\!p \; \umf_\sigma(\zeta)
                    \over      \bar{u}_{-\sigma}(k) \umf_\sigma(\zeta) }
    =  \sqrt{2}\; \sqrt{ {2\zeta p \over 2\zeta k} } s_\sigma(k,\hat{p}),
  \end{align}
\end{equation}
which also holds
in the general case of $u_\sigma(\beta)$, where
\begin{equation}
    \label{b-sigma2}
    b_\sigma(k,p)=
        {\sqrt{2} \over s_{-\sigma}(k,\beta)}
        \left( s_{-\sigma}(\beta,\hat{p}) s_\sigma(\hat{p},k) 
          +{m^2\over 2\zeta\hat{p}}\; \sqrt{ (2\beta\zeta)\; (2\zeta k) }
        \right).
\end{equation}

\section{\normalsize Born spin amplitudes}

Let us calculate lowest order spin amplitudes
for $e^-(p_1)e^+(p_2)\to f(p_3)\bar{f}(p_4)$.
For the moment we require $f\neq e$.
Using our basic massive spinors of eq.~(\ref{def-massive2}) 
with definite GPS helicities and Feynman rules, we define
\begin{equation}
  \begin{split}
    \Bmf \left[\hbox{}^{p}_{\lambda} \right] (X)
 =& \Bmf \left[ \hbox{}^{p_1}_{\lambda_1} \hbox{}^{p_2}_{\lambda_2}
                \hbox{}^{p_3}_{\lambda_3} \hbox{}^{p_4}_{\lambda_4} \right] (X) 
 =  ie^2 \sum_{B=\gamma,Z}
    \frac{ \bar{v}(p_2,\lambda_2) \gamma^\mu G^{e,B} u(p_1,\lambda_1)\;\;
           \bar{u}(p_3,\lambda_3) \gamma_\mu G^{f,B} v(p_4,\lambda_4)}
         { X^2 - {M_{B}}^2 +i\Gamma_{B} X^2 /M_{B} },
\\
      G^{e,B} =&\sum_{\lambda=\pm} \omega_\lambda g^{e,B}_\lambda,\quad
      G^{f,B} = \sum_{\lambda=\pm} \omega_\lambda g^{f,B}_\lambda,\\
  \end{split}
\end{equation}
where $g^{f,B}_\lambda$ are the usual chiral ($\lambda=+1,-1=R,L$) coupling constants
of the vector boson $B=\gamma,Z$ to fermion $f$ in units of the elementary charge $e$.

Spinor products are reorganized with the help of
the Chisholm identity (\ref{Chisholm}),
which applies assuming that electron spinors are massless,
and the inner product of eq.~(\ref{inner-massive}):
\begin{equation}
  \label{born}
    \Bmf \left[\hbox{}^{p}_{\lambda} \right] (X)
 = 2 ie^2  \sum_{B=\gamma,Z}
    \frac{ 
            \delta_{\lambda_1, -\lambda_2}
            \big[\;
                 g^{e,B}_{ \lambda_1} g^{f,B}_{-\lambda_1}\;
                 T_{ \lambda_3 \lambda_1}\; T'_{\lambda_2 \lambda_4}
                +g^{e,B}_{ \lambda_1} g^{f,B}_{ \lambda_1}\;
                 U'_{ \lambda_3 \lambda_2}\; U_{\lambda_1 \lambda_4}
            \big]
         }
         { X^2 - {M_{B}}^2 +i\Gamma_{B} X^2 /M_{B} },
\end{equation}
where
\begin{equation}
  \begin{split}
  T_{ \lambda_3 \lambda_1} =& \bar{u}(p_3, \lambda_3)  u(p_1, \lambda_1)
                            =S(p_3, m_3, \lambda_3,    p_1,   0,    \lambda_1),\\
  T'_{\lambda_2 \lambda_4} =& \bar{v}(p_2, \lambda_2)  v(p_4,  \lambda_4)
                            =S(p_2,   0,  -\lambda_2,  p_4,-m_4,   -\lambda_4 ),\\
  U'_{\lambda_3 \lambda_2} =& \bar{u}(p_3, \lambda_3)  v(p_2,-\lambda_2)
                            =S(p_3, m_3,\lambda_3,     p_2,   0,     \lambda_2),\\
  U_{ \lambda_1 \lambda_4} =& \bar{u}(p_1,-\lambda_1)  v(p_4, \lambda_4)
                            =S(p_1,   0,  -\lambda_1,  p_4,-m_4,    -\lambda_4).
  \end{split}
\end{equation}
We understand that the total $s$-channel four-momentum $X$ is always 
{\em the} four-vector that enters the $s$-channel vector boson propagators.
Let us stress that the above Born spin amplitudes will be used for $p_i$,
which {\em do not necessarily obey} the four-momentum conservation $p_1+p_2=p_3+p_4$.
This is necessary because, in the presence of the bremsstrahlung photons,
the relation $X=p_1+p_2=p_3+p_4$ does not usually hold.
Furthermore, any of the $p_i$ may, and occasionally will, be replaced by
the momentum $k$ of one of photons. In this case,
the spinor into which $k$ enters as an argument is understood to be massless.

\section{\normalsize First order, one virtual photon}

The \Order{\alpha^1} contribution with one virtual and zero real photon reads
\begin{equation}
\label{one-virtual}
    \Mcal^{(1)}_0 \left[\hbox{}^{p}_{\lambda} \right] (X)
  = \Bmf \left[\hbox{}^{p}_{\lambda} \right] (X) 
   \left[1+Q_e^2 F_1(s,m_\gamma) + Q_f^2 F_1(s,m_\gamma) \right]
   +\Mcal_{\rm box} \left[\hbox{}^{p}_{\lambda} \right] (X),
\end{equation}
where $F_1$ is the standard electric form-factor regularized with photon mass.
We omit, for the moment, the magnetic form-factor $F_2$; this is justified
for light final fermions. It will be restored in the future.
In $F_1$ we keep  the exact final fermion mass.

In the present work we use spin amplitudes for $\gamma$-$\gamma$ and $\gamma$-$Z$ boxes
in the small mass approximation $m_e^2/s\to 0, m_f^2/s\to 0$,
following refs.~\cite{was:1987,brown:1984},
\begin{equation}
  \label{boxy}
  \begin{split}
    \Mcal_{\rm Box} \left[\hbox{}^{p}_{\lambda} \right] (X)
& = 2 ie^2 \sum_{B=\gamma,Z}
    \frac{ 
                 g^{e,B}_{ \lambda_1} g^{f,B}_{-\lambda_1}\;
                 T_{ \lambda_3 \lambda_1} T'_{\lambda_2 \lambda_4}
                +g^{e,B}_{ \lambda_1} g^{f,B}_{ \lambda_1}\;
                 U'_{ \lambda_3 \lambda_2} U_{\lambda_1 \lambda_4}
         }
         { X^2 - {M_{B}}^2 +i\Gamma_{B} X^2 /M_{B} }\;
         \delta_{\lambda_1, -\lambda_2}
         \delta_{\lambda_3, -\lambda_4}\\
& \frac{\alpha}{\pi} Q_eQ_f
  \left[
         \delta_{\lambda_1,  \lambda_3}\;
         f_{\rm BDP}(\bar{M}^2_B,m_\gamma,s,t,u)
        -\delta_{\lambda_1, -\lambda_3}\;
         f_{\rm BDP}(\bar{M}^2_B,m_\gamma,s,u,t)
  \right],
  \end{split}
\end{equation}
where $\bar{M}^2_Z = M_Z^2-iM_Z\Gamma_Z$, $\bar{M}^2_\gamma=m_\gamma^2$,
and the function $f_{\rm BDP}$ is defined in eq.~(11) of ref.~\cite{brown:1984}.
The Mandelstam variables $s,t$ and $u$ are defined as usual.
Since in the rest of our calculation we do not use $m_f^2/s\to 0$,
we therefore intend to replace the above box spin amplitudes with
the finite-mass results. 
(NB: For the $\gamma$-$\gamma$ box the spin amplitudes with the exact final fermion mass%
\footnote{ It seems, however, that the $\gamma$-$Z$ box for 
  the heavy fermion is missing in the literature.}
were given in ref.~\cite{KORALB:1985}.)

\section{\normalsize First order 1-photon, ISR alone}

In order to introduce the notation gradually,
let us first consider the 1-photon emission matrix element separately for ISR.
The first order, 1-photon, ISR matrix element from the Feynman rule reads
\begin{equation}
  \label{isr-feynman}
  \begin{split}
    \Mcal^{\rm ISR}_1
    \left( \hbox{}^{p_1}_{\lambda_1} \hbox{}^{p_2}_{\lambda_2} \hbox{}^{k}_{\sigma}
    \right)=
   & {eQ_e \over 2kp_1}\;
    \bar{v}(p_2,\lambda_2)\; \mathbf{M}_1\;
         (\not\!{p_1}+m-{\not\!k})     \not\!{\epsilon}^\star_\sigma(k)\;
    u(p_1,\lambda_1)\\
   +&{eQ_e \over 2kp_2}\;
    \bar{v}(p_2,\lambda_2)
         \not\!{\epsilon}^\star_\sigma(k)\; (-{\not\!p_2}+m+\not\!{k}) \;
         \mathbf{M}_1\;
    u(p_1,\lambda_1),
  \end{split}
\end{equation}
where $\mathbf{M_1}$ is the annihilation scattering spinor matrix
(including final state spinors).
The above expression we split into soft IR parts%
\footnote{ This kind of separation was already  exploited
in refs.~\protect\cite{erw:1994}.
  We thank E. Richter-W\c{a}s for attracting our attention to this method.
  }
proportional to $(\not\!{p} \pm m)$
and non-IR parts proportional to ${\not\!k}$.
Employing the completeness relations of eq.~(\ref{transition-defs}) 
to those parts we obtain:
\begin{equation}
  \begin{split}
    \Mcal^{\rm ISR}_1
    \left( \hbox{}^{p_1}_{\lambda_1} \hbox{}^{p_2}_{\lambda_2} \hbox{}^{k}_{\sigma}
    \right)=
   &{eQ_e\over 2kp_1}\; \sum_\rho
     \Bmf_1\left[ \hbox{}^{p_1}_{\rho} \hbox{}^{p_2}_{\lambda_2}
           \right]
      U\left(     \hbox{}^k_\sigma     \right)\!
       \left[     \hbox{}^{p_1}_{\rho}
                  \hbox{}^{p_1}_{\lambda_1} \right]
    -{eQ_e\over 2kp_2}\; \sum_\rho
      V\left(   \hbox{}^k_\sigma      \right)\!
       \left[   \hbox{}^{p_2}_{\lambda_2}
                \hbox{}^{p_2}_{\rho}    \right]
      \Bmf_1 \left[ \hbox{}^{p_1}_{\lambda_1} \hbox{}^{p_2}_{\rho}
             \right]\\
   -&{eQ_e\over 2kp_1}\; \sum_\rho
     \Bmf_1\left[ \hbox{}^{k}_{\rho} \hbox{}^{p_2}_{\lambda_2}  \right]
      U\left(   \hbox{}^k_\sigma      \right)\!
       \left[   \hbox{}^{k}_{\rho}
                \hbox{}^{p_1}_{\lambda_1} \right]
    +{eQ_e\over 2kp_2}\; \sum_\rho
      V\left(   \hbox{}^k_\sigma      \right)\!
       \left[   \hbox{}^{p_2}_{\lambda_2}
                \hbox{}^{k}_{\rho}    \right]
      \Bmf_1 \left[ \hbox{}^{p_1}_{\lambda_1} \hbox{}^{k}_{\rho} \right],
  \end{split}
\end{equation}
where
$\Bmf_1 \left[ \hbox{}^{p_1}_{\lambda_1} \hbox{}^{p_2}_{\lambda_2} \right]
       =\bar{v}(p_2,\lambda_2) \mathbf{M}_1 u(p_1,\lambda_1).$
The summation in the first two terms gets eliminated due to the diagonality
property of $U$ and $V$, see eq.~(\ref{diagonality}), and leads to
\begin{equation}
  \label{first-order-isr}
  \begin{split}
   &\Mcal^{\rm ISR}_1
      \left( \hbox{}^{p_1}_{\lambda_1} \hbox{}^{p_2}_{\lambda_2} \hbox{}^{k}_{\sigma}
      \right)
   =\sfac^{(1)}_\sigma(k) 
    \Bmf_1  \left[ \hbox{}^{p_1}_{\lambda_1} \hbox{}^{p_2}_{\lambda_2} \right]
   +r^{(1)} \left[ \hbox{}^{p_1}_{\lambda_1} \hbox{}^{p_2}_{\lambda_2} \hbox{}^{k}_{\sigma} 
            \right](k),\\
  & r^{(1)} \left[ \hbox{}^{p_1}_{\lambda_1} \hbox{}^{p_2}_{\lambda_2}  \hbox{}^{k}_{\sigma} 
             \right](k)=
    -{eQ_e\over 2kp_1}\; \sum_\rho
     \Bmf_1\left[ \hbox{}^{k}_{\rho} \hbox{}^{p_2}_{\lambda_2}   \right]
        U\left(   \hbox{}^k_\sigma      \right)\!
         \left[   \hbox{}^{k}_{\rho} \hbox{}^{p_1}_{\lambda_1}   \right]
    +{eQ_e\over 2kp_2}\; \sum_\rho
        V\left(   \hbox{}^k_\sigma      \right)\!
         \left[   \hbox{}^{p_2}_{\lambda_2} \hbox{}^{k}_{\rho}   \right]
     \Bmf_1\left[ \hbox{}^{p_1}_{\lambda_1} \hbox{}^{k}_{\rho}   \right],\\
  & \sfac^{(1)}_\sigma(k) =
     eQ_e{b_\sigma(k,p_1) \over 2kp_1} -eQ_e{b_\sigma(k,p_2) \over 2kp_2},\quad
    |\sfac^{(1)}_\sigma(k)|^2 = 
             -\frac{e^2Q_e^2}{2} \bigg( {p_1\over kp_1} - {p_2\over k p_2}  \bigg)^2.
    \end{split}
\end{equation}
The soft part is now clearly separated and the remaining non-IR part,
necessary for the CEEX, is obtained.
The case of final state one real photon emission 
can be analysed in a similar way.


\section{\normalsize First order 1-photon ISR+FSR}

The first order, ISR+FSR, 1-photon matrix element, with explicit split
into IR and non-IR parts, reads
\begin{equation}
  \label{one-photon}
  \begin{split}
    \Mcal^{(1)}_1
      \left( \hbox{}^{p}_{\lambda} \hbox{}^{k}_{\sigma} \right)
  =
   \sfac^{(1)}_{\sigma}(k)\;
      \Bmf \left[ \hbox{}^{p}_{\lambda} \right] (P-k)
 + \sfac^{(0)}_{\sigma}(k)\;
      \Bmf  \left[ \hbox{}^{p}_{\lambda} \right] (P)
  +r^{(1)} \left[ \hbox{}^{p}_{\lambda} \hbox{}^k_\sigma  \right] (P-k)
  +r^{(0)} \left[ \hbox{}^{p}_{\lambda} \hbox{}^k_\sigma  \right] (P),
  \end{split}
\end{equation}
where we use the compact notation
$ \left[ \hbox{}^{p}_{\lambda} \right]
    \equiv  \left[ \hbox{}^{p_1}_{\lambda_1}
                   \hbox{}^{p_2}_{\lambda_2}
                   \hbox{}^{p_3}_{\lambda_3}
                   \hbox{}^{p_4}_{\lambda_4}
            \right],$
and the lowest order Born spin amplitudes $\Bmf$ are
defined in eq.~(\ref{born}).
The other ingredients are the initial state non-IR part:
\begin{equation}
  r^{(1)} \left[ \hbox{}^{p}_{\lambda} \hbox{}^k_\sigma \right] (X)
 ={ -eQ_e \over 2k p_1}
        \sum_\rho
         U    \left( \hbox{}^k_\sigma \right)\! 
              \left[ \hbox{}^{p_1}_{\lambda_1} \hbox{}^{k}_{\rho} \right]
         \Bmf \left[ \hbox{}^{k}_{\rho}
                     \hbox{}^{p_2}_{\lambda_2}
                     \hbox{}^{p_3}_{\lambda_3}
                     \hbox{}^{p_4}_{\lambda_4}
               \right] (X)
 +{ eQ_e \over 2k p_2}
        \sum_\rho
         V    \left( \hbox{}^k_\sigma \right)\!
              \left[ \hbox{}^{k}_{\rho} \hbox{}^{p_2}_{\lambda_2}\right]
         \Bmf \left[ \hbox{}^{p_1}_{\lambda_1}
                     \hbox{}^{k}_{\rho}
                     \hbox{}^{p_3}_{\lambda_3}
                     \hbox{}^{p_4}_{\lambda_4}
             \right] (X)
\end{equation}
and the final state non-IR part
\begin{equation}
  r^{(0)} \left[ \hbox{}^{p}_{\lambda} \hbox{}^k_\sigma \right]\! (X)
 =-{eQ_f \over 2k p_3}
        \sum_\rho
         U    \left( \hbox{}^k_\sigma \right)\!
              \left[ \hbox{}^{k}_{\rho} \hbox{}^{p_3}_{\lambda_3} \right]
         \Bmf \left[ \hbox{}^{p_1}_{\lambda_1}
                     \hbox{}^{p_2}_{\lambda_2}
                     \hbox{}^{k}_{\rho}
                     \hbox{}^{p_4}_{\lambda_4}
              \right] (X)
 +{ eQ_f \over 2k p_4}
       \sum_\rho
         V  \left( \hbox{}^k_\sigma \right)\!
            \left[ \hbox{}^{p_4}_{\lambda_4} \hbox{}^{k}_{\rho} \right]
         \Bmf
            \left[ \hbox{}^{p_1}_{\lambda_1}
                   \hbox{}^{p_2}_{\lambda_2}
                   \hbox{}^{p_3}_{\lambda_3}
                   \hbox{}^{k}_{\rho}
            \right] (X).
\end{equation}
The  FSR $\sfac$-factor
\begin{equation}
 \sfac^{(0)}_\sigma(k) =
            -eQ_f{b_\sigma(k,p_3) \over 2kp_3} +eQ_f{b_\sigma(k,p_4) \over 2kp_4},\quad
|\sfac^{(0)}_\sigma(k)|^2 = 
            -\frac{e^2Q_f^2}{2} \bigg( {p_3\over kp_3} - {p_4\over k p_4}  \bigg)^2
\end{equation}
we define analogously to the ISR case.

\section{\normalsize Coherent exclusive exponentiation, zero and first order}

Spin amplitudes in the zero-th order coherent exclusive exponentiation,
\Oceex{\alpha^0}, we define as follows 
\begin{equation}
  \label{M-expon0}
  \Meu^{(0)}_n 
    \left(   \hbox{}^{p}_{\lambda}
             \hbox{}^{k_1}_{\sigma_1} 
             \hbox{}^{k_2}_{\sigma_2}
             \dots
             \hbox{}^{k_n}_{\sigma_n}
    \right) =
    e^{\alpha B_4(p_1,...,p_4)}\;
    \sum_{\{\wp\}}\;
           { X^2_{\wp}  \over (p_3+p_4)^2 }
           \Bmf \left[  \hbox{}^{p}_{\lambda} \right]\!
           (X_\wp)\;\;
           \sfac_{\sigma_1}^{\wp_1}(k_1)
           \sfac_{\sigma_2}^{\wp_2}(k_2)
           \dots
           \sfac_{\sigma_n}^{\wp_n}(k_n),
\end{equation}
where the $s$-channel four-momentum in the resonance propagator is
$   X_\wp = p_1+p_2 -\sum_{i=1}^n   \wp_i\; k_i. $
The partition $\wp$ is defined as a vector $(\wp_1,\wp_2,\dots, \wp_n)$
where $\wp_i=1$ for ISR and  $\wp_i=0$ for FSR photon,
see the analogous construction in refs.~\cite{greco:1975,greco:1980}.
For a given partition $X_\wp$ is therefore the total incoming four-momentum
minus four-momenta of ISR photons.
The {\em coherent} sum is taken over set $\{ \wp \}$ of all $2^n$ partitions --
this set is explicitly the following
\begin{equation}
\{ \wp \} = \{
(0,0,0,\dots,0),\; (1,0,0,\dots,0),\;
(0,1,0,\dots,0),\; (1,1,0,\dots,0),\; \dots
(1,1,1,\dots,1) \}.
\end{equation}
In eq.~(\ref{M-expon0}) we profit from the
Yennie-Frautschi-Suura \cite{yfs:1961} fundamental proof of factorization
of all virtual IR corrections in the form-factor%
\footnote{ 
  In the LL approximation it is, of course, the doubly-logarithmic Sudakov form-factor.}
$\exp(\alpha B_4)$, where
\begin{equation}
  \begin{split}
   &B_4(p_1,...,p_4) 
  =  Q_e^2   B_2(p_1,p_2)  +Q_f^2   B_2(p_3,p_4)\\
&\qquad\qquad
    +Q_e Q_f B_2(p_1,p_3)  +Q_e Q_f B_2(p_2,p_4) 
    -Q_e Q_f B_2(p_1,p_4)  -Q_e Q_f B_2(p_2,p_3).
\\& B_2(p,q) \equiv
    \int {d^4k\over k^2 -m_\gamma^2 +i\epsilon}\; {i\over (2\pi)^3}\;
             \bigg( {2p+k \over k^2 +2kp +i\epsilon} 
                   +{2q-k \over k^2 -2kq +i\epsilon}  \bigg)^2.
  \end{split}
\end{equation}
In the above we assume that IR singularities are regularized with a finite
photon mass $m_\gamma$ which enters into all $B_2$'s and implicitly into $\sfac$-factors 
(and the real photon phase space integrals, in the following discussion).

The auxiliary factor  $F = X^2 / (p_3+p_4)^2$ is,
from the formal point of view, not really necessary.
Note that the $F$-factor does not affect the soft limit; it really matters
if at least one very hard FSR photon is present.
However, the $F$-factor is very useful, because it is present in the
photon emission matrix element, both in \Order{\alpha^1} and also in all orders
in the leading logarithmic (LL) approximation.
It has also been present for a long time now in the ``crude distribution'' 
in the YFS-type Monte Carlo generators, see for instance ref.~\cite{yfs2:1990}.
It is therefore natural to include it already in the \Order{\alpha^0} exponentiation.
Otherwise, this $F$-factor will be included order by order.
However, in such a case, the convergence of perturbative expansion will be deteriorated.
As we shall see below, the introduction of the $F$-factor will slightly complicate
the first order exponentiation.

The complete set of spin amplitudes
for emission of $n$ photons we define in \Oceex{\alpha^1} as follows:
\begin{equation}
  \label{M-expon1}
  \begin{split}
 &\Meu^{(1)}_n \left( \hbox{}^{p}_{\lambda}
                      \hbox{}^{k_1}_{\sigma_1} 
                      \hbox{}^{k_2}_{\sigma_2}
                      \dots
                      \hbox{}^{k_n}_{\sigma_n}
               \right)
   =e^{\alpha B_4(p_1,...,p_4)}\\
&\qquad\times
    \sum_{\{\wp\}}\;  
    \prod_{i=1}^n \;
           \sfac_{\sigma_i}^{\wp_i}(k_i)\;
    \bigg(
           \Bmf \left[  \hbox{}^{p}_{\lambda} \right] (X_\wp)
           \left(1+ \delta^{(1)}_{Virt}\right)
           +\Reu_{\rm Box} \left[\hbox{}^{p}_{\lambda} \right] (X_\wp)
           +\sum_{j=1}^n 
            \Reu^{(\wp_j)}_1
               \left[ \hbox{}^{p}_{\lambda} \hbox{}^{k_j}_{\sigma_j} \right] (X_\wp)
    \bigg),\\
 &\Reu^{(\omega)}_1  \left[ \hbox{}^{p}_{\lambda} \hbox{}^{k}_{\sigma} \right] (X)
    \equiv 
     {1 \over \sfac_{\sigma}^{\omega}(k) }\!
     \left[
     r^{(\omega)} \left[ \hbox{}^{p}_{\lambda} \hbox{}^{k}_{\sigma} \right]\! (X)\;
     +\left( { (p_3+p_4+\omega k_j)^2  \over (p_3+p_4)^2} - 1 \right)
      \Bmf \left[  \hbox{}^{p}_{\lambda} \right]\! (X),\;\; \omega=\pm 1
    \right].
  \end{split}
\end{equation}
The IR-finite $\delta^{(1)}_{Virt}$ and $\Reu_{\rm Box}$ are determined 
{\em unambiguously}
by identifying for $n=0$ the above equation with eq.~(\ref{one-virtual}),
up to terms of \Order{\alpha^1}. We obtain
\begin{equation}
  \delta^{(1)}_{Virt}(s) = 
   Q_e^2 F_1(s,m_\gamma) + Q_f^2 F_1(s,m_\gamma)
  -Q_e^2 \alpha B_2(s,m_\gamma) -Q_f^2 \alpha B_2(s,m_\gamma).
\end{equation}
The $\Reu_{\rm Box}$ is obtained from $\Mcal_{\rm Box}$
by means of the substitution%
\footnote{
  In the above procedure of subtracting IR divergences,
  there is no reference to cut on photon energy, only reference to the photon mass,
  similar to the YFS exponentiation on squared spin-summed amplitudes.}
\begin{equation}
 f_{\rm BDP}(\bar{M}^2_B,m_\gamma,s,t,u) 
   \to
 f_{\rm BDP}(\bar{M}^2_B,m_\gamma,s,t,u)  - f_{\rm IR}(m_\gamma,t,u),
\end{equation}
where
\begin{equation}
    f_{\rm IR}(m_\gamma,t,u)
          = \frac{2}{\pi} B_2(m_\gamma,t) -\frac{2}{\pi} B_2(m_\gamma,u)
          = \ln\left( \frac{t}{u} \right) 
            \ln\left( \frac{m_\gamma^2}{\sqrt{tu}} \right)
           +\frac{1}{2} \ln\left( \frac{t}{u} \right).
\end{equation}
Similarly the IR-finite $\Reu^{(\omega)}_1$ is determined {\em uniquely}
by identifying, for $n=1$, eq.~(\ref{M-expon1}) with eq.~(\ref{one-photon}).
In particular the factor $F-1=(p_3+p_4+\omega k_j)^2/(p_3+p_4)^2 - 1$ is a consequence of
the introduction of the $F$-factor in eq.~(\ref{M-expon0}).
If it was not included, then the 1-photon part in eq.~(\ref{M-expon1})
would not reduce to the amplitude of eq.~(\ref{one-photon}).
Thanks to the presence of $F-1$,  for $n=1$, we recover in eq.~(\ref{M-expon1})
the correct first order amplitude of eq.~(\ref{first-order-isr}).

For very narrow resonances the photon emission in the decay process
is separated from the photon emission in the production process by
very large time-space distance.
The ISR*FSR interference is therefore strongly suppressed, typically by $\Gamma/M$ factors.
Since our real photons are present down to arbitrarily low
$k^0_{\min} =\epsilon \sqrt{s}/2 \ll \Gamma $, 
the effects due to the resonance complex phase in the emission of the
{\em real} photons are taken into account {\em numerically} and exactly.
For {\em virtual} photons we have to sum up {\em analytically}
certain subset of the ISR*FSR interferences
to infinite order following Greco et al.~\cite{greco:1975,greco:1980}.
In practice the rule is:
multiply each part of the spin amplitude
proportional to Z-propagator by the additional factor $\exp(\delta_G(s,t,u))$
where:
\begin{equation}
\delta_G(s,t,u)  = -2Q_e Q_f {\alpha\over\pi} 
                    \ln\left( {t\over u} \right) 
                    \ln\left( {M_Z^2-iM_Z \Gamma_Z -s \over M_Z^2} \right)
\end{equation}
In \Order{\alpha^1} the above exponential factor induces the additional subtraction
in the $\gamma$-Z box:
$ \Meu_{\rm box}(s,t,u) \to \Meu_{\rm box}(s,t,u) - \delta_G(s,t,u)$.
Strictly speaking the above improvement is not really necessary,
because we would have obtained it order-by-order, through higher order virtual non-IR correction.
In practice, however, it is mandatory. If we had not made it, then
the  ISR*FSR interference contribution to $A_{FB}$ at Z peak from \Oceex{\alpha^1}
would be dramatically wrong, i.e. 0.5\% instead of 0.05\%!

\section{\normalsize Differential cross sections and the YFS form-factor}

The master formula for the unpolarized \Oceex{\alpha^r}
total cross section is given by the standard quantum-mechanical expression
of the type ``matrix element squared modulus times phase space''
(contrary to typical ``parton shower'' approach)
\begin{equation}
\label{eq:master}
  \sigma^{(r)} = 
  \sum_{n=0}^\infty {1\over n!}
  \int d\tau_{n} ( p_1+p_2 ;\; p_3,p_4,\; k_1,\dots,k_n)\;
  \frac{1}{4}
  \sum_{\lambda,\sigma_1,\dots,\sigma_n =\pm}\;
  \left|
    \Meu^{(r)}_n 
    \left(  \hbox{}^{p}_{\lambda}
            \hbox{}^{k_1}_{\sigma_1} 
            \hbox{}^{k_2}_{\sigma_2}
            \dots
            \hbox{}^{k_n}_{\sigma_n}
    \right)
  \right|^2,
\end{equation}
where the Lorentz invariant phase space (LIPS) integration element is
\begin{equation}
  \int d\tau_n(P;p_1,p_2,...p_n) \equiv
  \int (2\pi)^4  \delta^{(4)}\left( P -\sum_{j=1}^n p_j \right)
  \prod_{j=1}^n  {d^3 p_j\over 2p^0_j (2\pi)^3}.
\end{equation}
The above total cross section is perfectly
IR-finite, as can be checked with a little bit of effort
by {\em analytical} partial differentiation%
\footnote{ This method of validating IR-finiteness was noticed by G. Burgers~\cite{burgers}.
  The classical method of ref.~\protect\cite{yfs:1961} relies on the techniques of the
  Melin transform.}
with respect the photon mass
\begin{equation}
  \frac{\partial}{\partial m_\gamma} \sigma^{(r)} =0.
\end{equation}
Furthermore, the integral of eq.~(\ref{eq:master}) is perfectly implementable
in the Monte Carlo form,
using a method very similar to those in ref.~\cite{yfs2:1990}.
Traditionally, however, the lower boundary on the real soft photons is defined
using the energy cut condition $k^0>\varepsilon \sqrt{s}/2$ in the laboratory frame.
The practical advantage of such a cut is the lower photon multiplicity in the MC simulation,
and consequently a faster computer program%
\footnote{ The disadvantage of the cut $k^0>\varepsilon \sqrt{s}/2$ is that in the MC 
  it has to be implemented in {\em different} reference frames for ISR and for FSR -- this costs
  the additional delicate procedure of bringing these two boundaries together,
  see ref.~\cite{KK:1998} and/or discussion in the analogous 
  $t$-channel case in ref.~\cite{bhlumi2:1992}.}.
If the above energy cut on the photon energy is adopted, then the real soft-photon
integral between the lower LIPS boundary defined by $m_\gamma$ and that
defined by $\varepsilon$ can be evaluated by hand and summed up rigorously
(the only approximation is $m_\gamma/m_e \to 0$)
into an additional overall factor $\exp(2\alpha \tilde{B}_4(p_1,...,p_4))$, where
\begin{equation}
  \begin{split}
   &\tilde{B}_4(p_1,...,p_4) 
  =  Q_e^2   \tilde{B}_2(p_1,p_2)  +Q_f^2   \tilde{B}_2(p_3,p_4)\\
&\qquad\qquad
    +Q_e Q_f \tilde{B}_2(p_1,p_3)  +Q_e Q_f \tilde{B}_2(p_2,p_4) 
    -Q_e Q_f \tilde{B}_2(p_1,p_4)  -Q_e Q_f \tilde{B}_2(p_2,p_3),
\\& \tilde{B}_2(p,q) \equiv
             \int\limits_{k^0<\varepsilon \sqrt{s}/2} 
             {d^3k\over k^0}\;
             \frac{(-1)}{8\pi^2} \bigg( {p\over kp} - {q\over k q}  \bigg)^2.
  \end{split}
\end{equation}
Let us introduce $\Mmf^{(r)}_n =e^{-\alpha B_4} \Meu^{(r)}_n$ 
(without virtual IR singularities) and,
altogether, the above reorganization yields the new expression for the 
unpolarized total cross-section
\begin{equation}
\label{eq:master2}
\begin{split}
  \sigma^{(r)} = &
  \sum_{n=0}^\infty {1\over n!}
  \int d\tau_{n} ( p_1+p_2 ;\; p_3,p_4,\; k_1,\dots,k_n)\;
  e^{Y(p_1,...,p_4)}\;
  \frac{1}{4}
  \sum_{\lambda,\sigma_i=\pm}\;
  \left|
    \Mmf^{(r)}_n 
    \left(  \hbox{}^{p}_{\lambda}
            \hbox{}^{k_1}_{\sigma_1} 
            \hbox{}^{k_2}_{\sigma_2}
            \dots
            \hbox{}^{k_n}_{\sigma_n}
    \right)
 \right|^2
\end{split}
\end{equation}
where
$Y(p_1,...,p_4) = 2\alpha \tilde{B}_4(p_1,...,p_4) + 2\alpha \Re B_4(p_1,...,p_4)$
is the conventional YFS form-factor defined analytically in terms of logs and Spence
functions -- we do not show it here explicitly due to lack of space,
see refs.~\cite{bhwide:1997,yfsww:1996,yfsww:1998,KK:1998}.
In the YFS form-factors we keep the final fermion mass exact.
The fully exclusive differential cross section of eq.~(\ref{eq:master2}) 
is already implemented
in the Monte Carlo event generator ${\cal KK}$ \cite{KK:1998}.

The extension of the above exponentiation procedure to \Oceex{\alpha^2}
and beyond requires more work, but does not pose any conceptual problem.
It will be implemented in the future version of the ${\cal KK}$ Monte Carlo.

\section{\normalsize Fermion spin polarization and photon spin randomization}

The great advantage of working with spin amplitudes is the easiness
of introduction of full spin polarizations for all particles.
The general case of the total cross section with polarized beams
and decays of unstable final fermion being sensitive to spin polarization
\cite{jadach_was:1984,tauola2.4,jadach:1985,gps:1998} reads
\begin{equation}
  \label{sigma-polarized}
  \begin{split}
  \sigma^{(r)} = &
  \sum_{n=0}^\infty {1\over n!}
  \int d\tau_{n} ( p_1+p_2 ;\; p_3,p_4,\; k_1,\dots,k_n)\;
  e^{Y(p_1,...,p_4)}\;
    \sum_{\sigma_i}\;
    \sum_{a,b,c,d=0}^3\;
    \sum_{\lambda_i,\bar{\lambda}_i}
\\&\quad
        \hat{\varepsilon}^a_1 \hat{\varepsilon}^b_2\;
        \sigma^a_{\lambda_1 \bar{\lambda}_1}
        \sigma^b_{\lambda_2 \bar{\lambda}_2}
    \Mmf^{(r)}_n 
    \left(  \hbox{}^{p}_{\lambda}
            \hbox{}^{k_1}_{\sigma_1} 
            \hbox{}^{k_2}_{\sigma_2}
            \dots
            \hbox{}^{k_n}_{\sigma_n}
    \right)
    \left[
    \Mmf^{(r)}_n 
    \left(  \hbox{}^{p}_{\bar{\lambda}}
            \hbox{}^{k_1}_{\sigma_1} 
            \hbox{}^{k_2}_{\sigma_2}
            \dots
            \hbox{}^{k_n}_{\sigma_n}
    \right)
    \right]^\star
        \sigma^c_{\bar{\lambda}_3 \lambda_3 }
        \sigma^d_{\bar{\lambda}_4 \lambda_4 }
        \hat{h}^c_3 \hat{h}^d_4,
  \end{split}
\end{equation}
where, for $k=1,2,3$,  $\sigma^k$ are Pauli matrices and
$\sigma^0_{\lambda,\mu} = \delta_{\lambda,\mu}$ is the unit matrix.
The components $\hat{\varepsilon}^a_1, \hat{\varepsilon}^b_2, a,b=1,2,3$ 
are the components of the conventional spin polarization vectors 
of $e^-$ and $e^+$ respectively, defined in the so-called GPS fermion rest frames
(see ref~\cite{gps:1998} for the exact definition of these frames).
We define $\hat{\varepsilon}^0_i=1$
in a non-standard way  (i.e. $p_i\cdot \hat{\varepsilon}_i=m_e$).
The {\em polarimeter} vectors $\hat{h}_i$ are similarly defined
in the appropriate GPS rest frames of the final unstable fermions  ($p_i\cdot \hat{h}_i=m_f$).
Note that, in general, $\hat{h}_i$ may
depend in a non-trivial way on momenta of all decay products,
see refs.~\cite{jadach:1985,tauola2.4} for details.
We did not introduce polarimeter vectors for bremsstrahlung photons,
i.e. we take advantage of the fact that the high energy experiment is completely blind
to photon spin polarizations.

Let us finally touch briefly upon one very serious problem and its solution.
In eq.~(\ref{sigma-polarized})  the single spin amplitude $\Mmf^{(1)}_n$ already contains
$2^n(n+1)$ terms (due to $2^n$ ISR--FSR partitions).
The grand sum over spins in eq.~(\ref{sigma-polarized}) counts
$2^n 4^4 4^4 = 2^{n+16}$ terms! 
Altogether we expect up to $N\sim n2^{2n+16}$ operations
in the CPU time expensive complex (16 bytes) arithmetics.
Typically in $e^-e^+\to \mu^-\mu^+$ the average photon multiplicity with $k^0>1$MeV
is about 3, corresponding to $N\sim 10^7$ terms.
In a sample of $10^4$ MC events there will be a couple of events with 
$n=10$ and $N=10^{12}$ terms, clearly something that would ``choke''
completely any modern, fast workstation.
There are several simple tricks that help to soften the problem;
for instance, objects such as
$\sum_a \hat{\varepsilon}^a_i \sigma^a_{\lambda\bar{\lambda}}$ and the $\sfac$-factors
are evaluated only once and stored for multiple use.
This is however not sufficient. 
What really helps to substantially speed up the numerical calculation
in the Monte Carlo program is the following trick of {\em photon spin randomization}.
Instead of evaluating
the sum over photon spins $\sigma_i,\;i=1,...,n$  in eq.~(\ref{sigma-polarized}),
we generate randomly one spin sequence of $(\sigma_1,...,\sigma_n)$ per MC event
and the MC weight is calculated only for this particular spin sequence!
In this way we save one hefty $2^n$ factor in the calculation time%
\footnote{The other $2^n$ factor due to coherent summation over partitions cannot
  be eliminated, unless we give up on narrow resonances.}.
Mathematically this method is correct, i.e.
the resulting cross section and all MC distribution
will be the same as if we had used in the MC weight 
the original  eq.~(\ref{sigma-polarized})
(see a formal proof of the above statement in Sect.~4 of ref.~\cite{MCguide}).
Let us stress again that it is possible to apply this photon spin randomization
trick because
(a) the typical high energy experiment is blind to photon spin polarization, so that
we did not need to introduce in eq.~(\ref{sigma-polarized}) 
the polarimeter vectors for photons, and
(b) For our choice of photon spin polarizations 
the cross section is rather weakly sensitive to them,
so the method does not lead to significant loss in the MC efficiency.

\section{\normalsize Conclusions}
We presented the first order coherent exclusive exponentiation CEEX scheme,
with the full control over spin polarization for all fermions.
This new method of exponentiation is very general and
has many immediate and longer term advantages.
The immediate profit will be the inclusion of the ISR--FSR interferences 
and availability of the exact distributions for multiple hard photons
without giving up on exclusive, YFS-style, exponentiation.
In particular it is applicable to diffucult case of the narrow resonances.
The resulting spin amplitudes and the differential distributions are readily implemented
in the MC event generator. (Numerical results will be presented elsewhere.)

\vspace{4mm}
\noindent
{\bf Acknowledgements}\\
We thank the CERN Theory Division and all four LEP collaborations for support.
One of us (Z.W.) acknowledges specially the support of the ETH L3 group during the 
final work on the paper preparation.
Useful discussions with E.~Richter-W\c{a}s are warmly acknowledged.
We also thank W. P\l{}aczek for correcting the manuscript.

\bibliographystyle{prsty}
\bibliography{KK}

\end{document}